\documentclass[sigconf,authorversion,screen]{aamas}

\usepackage{balance} 
\usepackage{subcaption}
\usepackage{algorithm}
\usepackage{algpseudocode}
\usepackage{float}

\doi{CYXP1261}

\makeatletter
\gdef\@copyrightpermission{
  \begin{minipage}{0.2\columnwidth}
   \href{https://creativecommons.org/licenses/by/4.0/}{\includegraphics[width=0.90\textwidth]{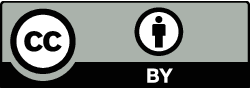}}
  \end{minipage}\hfill
  \begin{minipage}{0.8\columnwidth}
   \href{https://creativecommons.org/licenses/by/4.0/}{This work is licensed under a Creative Commons Attribution International 4.0 License.}
  \end{minipage}
  \vspace{5pt}
}
\makeatother

\setcopyright{ifaamas}
\acmConference[AAMAS '26]{Proc.\@ of the 25th International Conference
on Autonomous Agents and Multiagent Systems (AAMAS 2026)}{May 25 -- 29, 2026}
{Paphos, Cyprus}{C.~Amato, L.~Dennis, V.~Mascardi, J.~Thangarajah (eds.)}
\copyrightyear{2026}
\acmYear{2026}
\acmDOI{}
\acmPrice{}
\acmISBN{}

\acmSubmissionID{1589}

\title{Too Many Specialists: Emergent Inefficiencies and Bottlenecks for Multi-agent Ad-hoc Collaboration}
\subtitle{Extended Abstract}  
\author{Benjamin Panny}
\affiliation{
  \institution{University of Pittsburgh}
  \city{Pittsburgh, PA}
  \country{United States}}
\email{BMP83@pitt.edu}

\author{Shashank Mehrotra}
\affiliation{
  \institution{Honda Research Institute USA, Inc.}
  \city{San Jose, CA}
  \country{United States}}
\email{shashank\_mehrotra@honda-ri.com}

\author{Zahra Zahedi}
\affiliation{
  \institution{Honda Research Institute USA, Inc.}
  \city{San Jose, CA}
  \country{United States}}
\email{zahra\_zahedi@honda-ri.com}

\author{Teruhisa Misu}
\affiliation{
  \institution{Honda Research Institute USA, Inc.}
  \city{San Jose, CA}
  \country{United States}}
\email{tmisu@honda-ri.com}

\author{Kumar Akash}
\affiliation{
  \institution{Honda Research Institute USA, Inc.}
  \city{San Jose, CA}
  \country{United States}}
\email{kakash@honda-ri.com}

\begin{abstract}
Computational models of collaboration without prior coordination often overlook how heterogeneous agent traits and complex task structures jointly produce systemic bottlenecks, inefficiencies, and contribution inequalities. We address this by using an agent-based model of ad-hoc teamwork in a kitchen environment. Our model integrates diverse agent personas with tasks that combine serial and parallel dependencies.  
We identify a \textit{specialist's dilemma}, where rigid role assertion generates system-level bottlenecks, amplifies workload inequality, and fosters fragmented, homophilous networks. We also find that team size and communication overhead interact with problem structure to generate diminishing returns and redundant collaboration. Linking micro-level behavior to macro-level outcomes provides insights into emergent collaboration and design principles for effective multi-agent teamwork.
\end{abstract}


\newcommand{\BibTeX}{\rm B\kern-.05em{\sc i\kern-.025em b}\kern-.08em\TeX}

\begin{document}


\pagestyle{fancy}
\fancyhead{}


\maketitle


\section{Introduction}
Collaboration requires agents to balance initiative, adaptability, and mutual awareness under uncertainty  \cite{tambe1997towards}, particularly in ad-hoc settings where agents lack prior coordination. Computational models of teamwork can alleviate these challenges by predicting the conditions under which collaboration will succeed or fail \cite{salas_is_2005,spain2019toward}.

Frameworks such as the Transactive Systems framework \cite{wegner_transactive_1987,ickes_cognitive_1985,ren_transactive_2011, gupta_articulating_2021} and Collective Adaptation \cite{galesic_beyond_2023} highlight how socially networked reasoning and knowledge enable team reconfiguration when environments change. However, current models rarely scale beyond small teams or account for serial-parallel task structures \cite{lim_kill_2023, mason_collaborative_2012}. As a result, few models formally link macro-level performance, bottlenecks, and inequalities observed in real teams to micro-level agent behaviors observed in the individuals that constitute these teams.

In this paper, we address this gap through an agent-based model of ad-hoc teamwork in a physically grounded kitchen environment. By introducing heterogeneous agent personas endowed with social traits---such as collaboration initiative, agreeableness, and skill assertion---we show how bottlenecks, inequalities, and subgroup formations can emerge from simple local interaction rules. Our findings offer mechanistic levers for human-AI team design: seeding proactive leadership, mitigating rigid specialization, and bridging silos through adaptive coordination. 

\begin{figure}[t]
 \centering
 \includegraphics[width=1\linewidth]{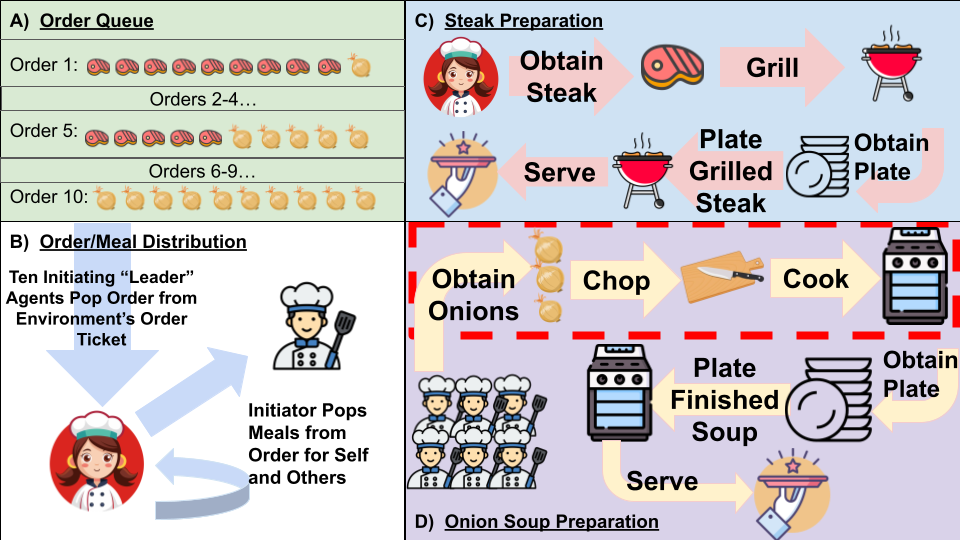}
 \caption{Kitchen Environment and Problem Structure. \textbf{(A)} Orders vary the ratios of parallel (Soup) vs. serial (Steak) meals. \textbf{(B)} Agents claiming tasks become ``leaders.'' \textbf{(C)} Serial Steak sequence. \textbf{(D)} Parallelizable Soup cycle.}.
 \label{fig:problem_structure}
 \Description{Kitchen Environment Problem Structure. A) Ten orders of ten meals each, varying the ratio of Onion Soup (parallel) to Steak (serial). B) Agents claiming orders/meals become "leaders," distributing sub-tasks to peers. C) Serial steps for Steak. D) Onion Soup steps; the red box marks the parallelizable cycle.}
 \vspace{-1em}
\end{figure}

\section{Collaboration Model}

We developed an agent-based model (ABM) where agents collaborate to complete cooking tasks in a 2D grid environment (Cf. \cite{carroll2020overcooked}). We vary team size \cite{lariviere_team_2015}, communication costs \cite{odaniel_professional_2008}, and agent social/transactive factors \cite{gupta_transactive_2022, galesic_beyond_2023} (Figure \ref{fig:problem_structure}).

\subsection{Serial and Parallel Kitchen Tasks}
Real-world collaboration tasks are rarely perfectly divisible. We designed two recipes modeling this constraint: \textbf{ (1) Steak:} A serial task with a fixed sequence (get meat, grill, plate, serve). \textbf{(2) Onion Soup:} A parallelizable task with repeated substeps (get onion, chop onion, place in stove) that multiple agents can perform concurrently. 

\subsection{Agent Attributes Determine Agent Rule-sets}
Agents' attributes result in actions via a decentralized Affordance-Context-Action (ACA) loop governed by four randomly assigned traits. \textbf{Agreeableness} controls the likelihood of accepting help requests, enabling tests of passive cooperation. \textbf{Collaboration Initiative} determines whether an agent initiates interactions, allowing us to measure the impact of proactive communicators. \textbf{Task Distribution Preference} governs the choice between starting unclaimed tasks or joining existing ones, introducing differences in organizational strategy (i.e., complete started tasks before starting new tasks or not). Finally, \textbf{Skill Assertion} determines whether agents refuse tasks outside their expertise; this operationalizes the \emph{specialist's dilemma}~\cite{olson1965logic, gupta_transactive_2022} by introducing tension between individual competence and team efficiency.
\section{Results}
\subsection{Scaling Effects and Diminishing Returns}
\begin{figure}[h]
    \centering
    \includegraphics[width=0.6\linewidth]{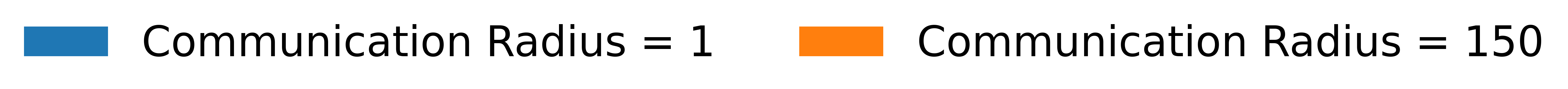}\\
    
    \begin{subfigure}[b]{0.48\linewidth}
        \centering
        \includegraphics[width=\textwidth]{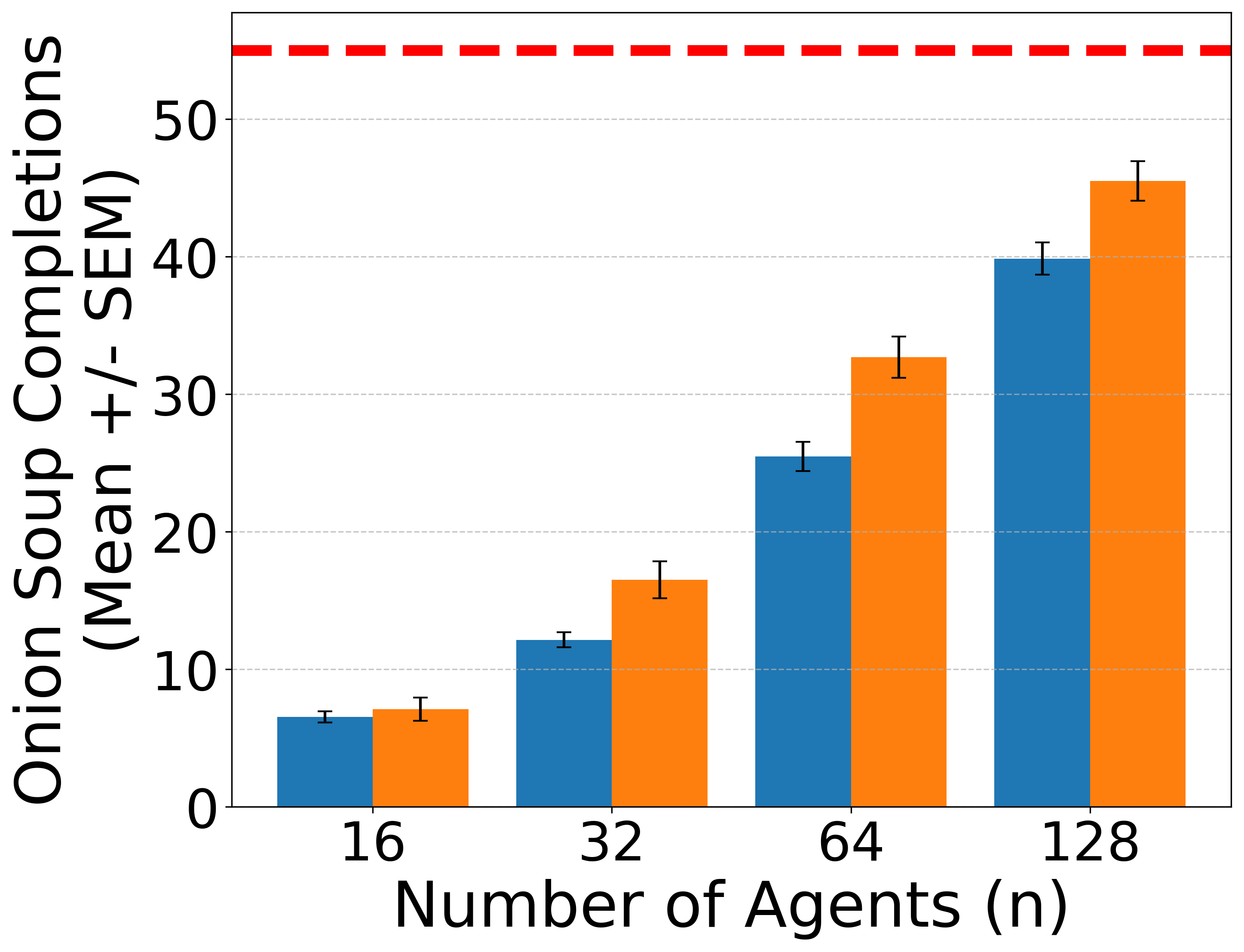}
        \caption{Onion Soup (Parallel)}
        \label{fig:onion_soup_bar_n_comm}
    \end{subfigure}
    \hfill
    \begin{subfigure}[b]{0.48\linewidth}
        \centering
        \includegraphics[width=\textwidth]{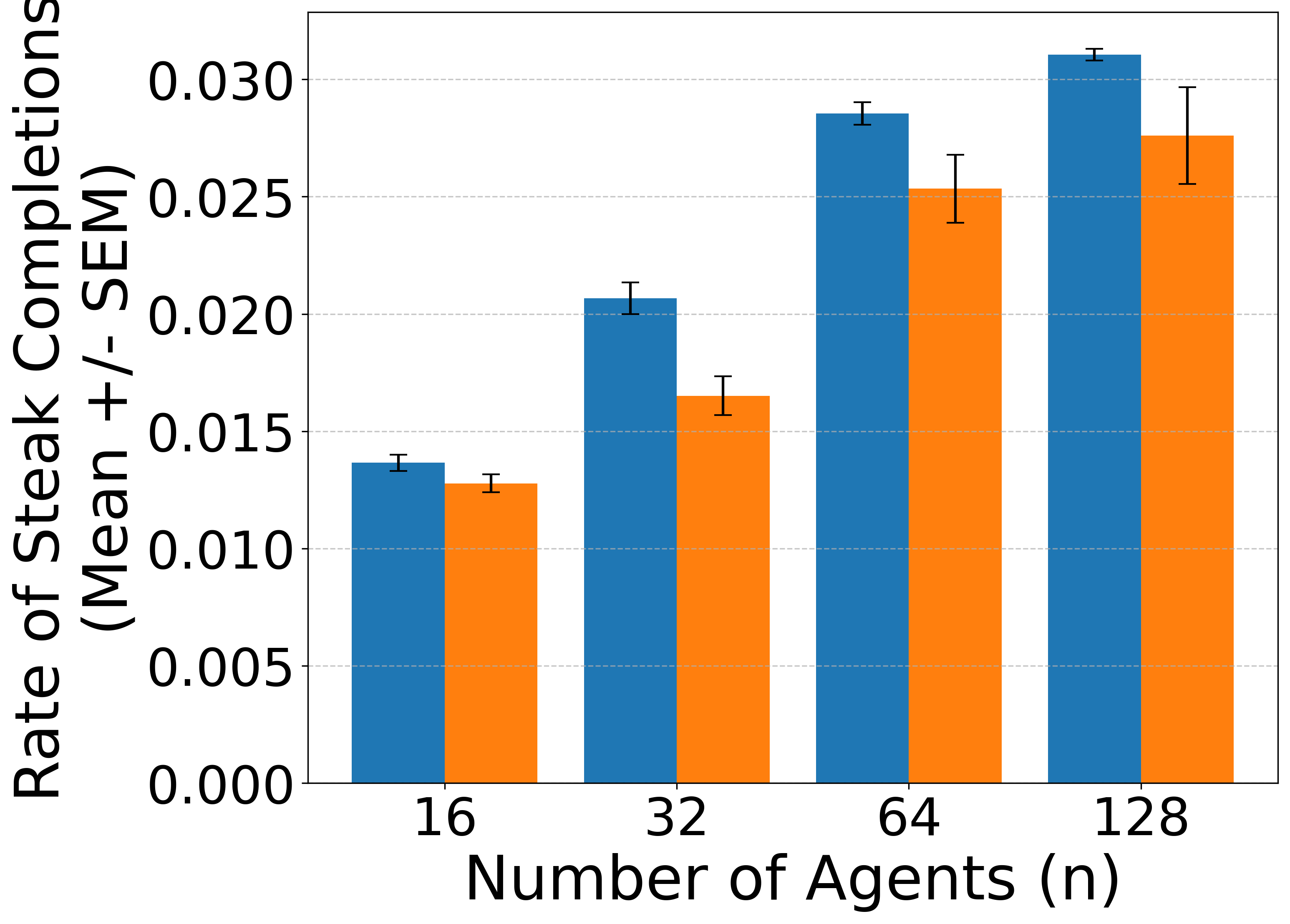}
        \caption{Steak (Serial)}
        \label{fig:steak_slope_barplot_n_comm}
    \end{subfigure}

    \caption{\textbf{(a)} Parallel tasks (Onion Soup) scale with team size and broad communication. \textbf{(b)} Serial tasks (Steak) benefit from higher communication cost, which results in smaller, more efficient teams.}
    \Description{Parallel tasks (Onion Soup) scale with team size and broad communication. Serial tasks (Steak) benefit from restricted communication, which minimizes overhead by enforcing smaller, more efficient teams.}
\end{figure}

Team size and communication affected serial and parallel tasks differently. For the parallelizable onion soup task (Figure \ref{fig:onion_soup_bar_n_comm}), larger teams and lower communication costs produced significantly more meals, though this was subject to diminishing returns. For the serial steak task, larger teams and \textit{higher} communication costs produced steaks faster than their counterparts with low communication costs (Figure \ref{fig:steak_slope_barplot_n_comm}). This is because low communication costs promoted larger teams that were effective for parallel tasks, but redundant for serial tasks. Thus, coordination overhead can be beneficial when it restricts redundant team building.

\subsection{Bottlenecks from Specialization}
We identified a \textit{specialist's dilemma}, where agents asserting their specific skills creates system-level bottlenecks.

A high proportion (100\%) of skill-asserting agents resulted in a significant decrease in total meals completed. However, this was partially rectified when specialists were also proactive communicators with task distribution preferences, as this persona enabled specialists to find tasks that fit their specialty earlier in the simulation.

Skill assertion also emerged as the primary driver of workload imbalance. When agents rigidly adhered to roles (e.g., "I only grill"), they remained idle when they could have supported agents working on other tasks.

\subsection{Emergent Silos}
\begin{figure}[H]
    \centering
    \includegraphics[width=0.9\columnwidth]{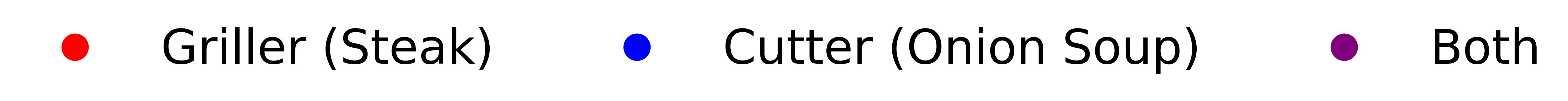}\vspace{-5pt}
     \\
    \begin{subfigure}[b]{0.48\columnwidth}
        \centering
        \includegraphics[width=\textwidth]{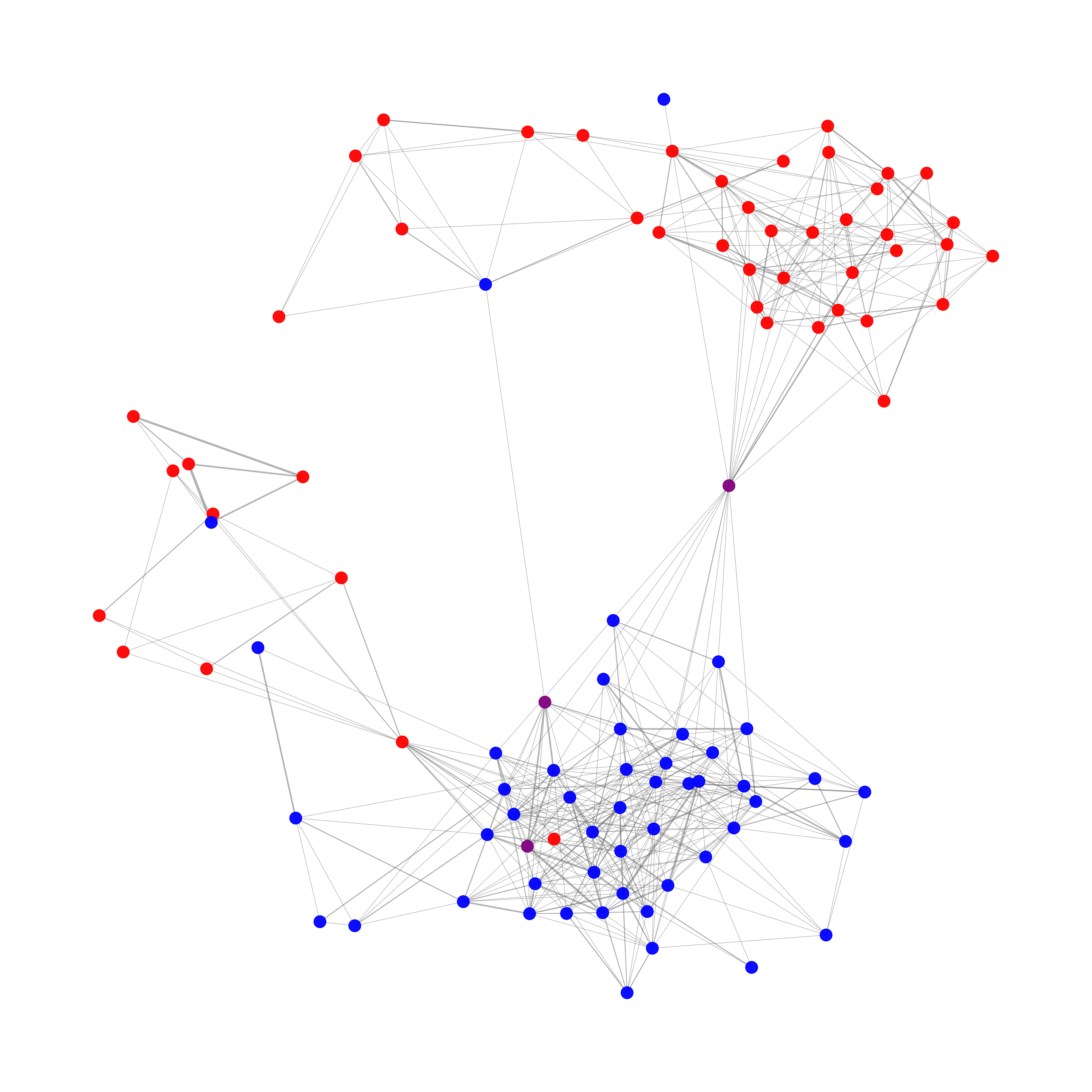}
        \caption{100\% Skill Assertion}
        \label{fig:homophily_sa1}
    \end{subfigure}
    \hfill
    \begin{subfigure}[b]{0.48\columnwidth}
        \centering
        \includegraphics[width=\textwidth]{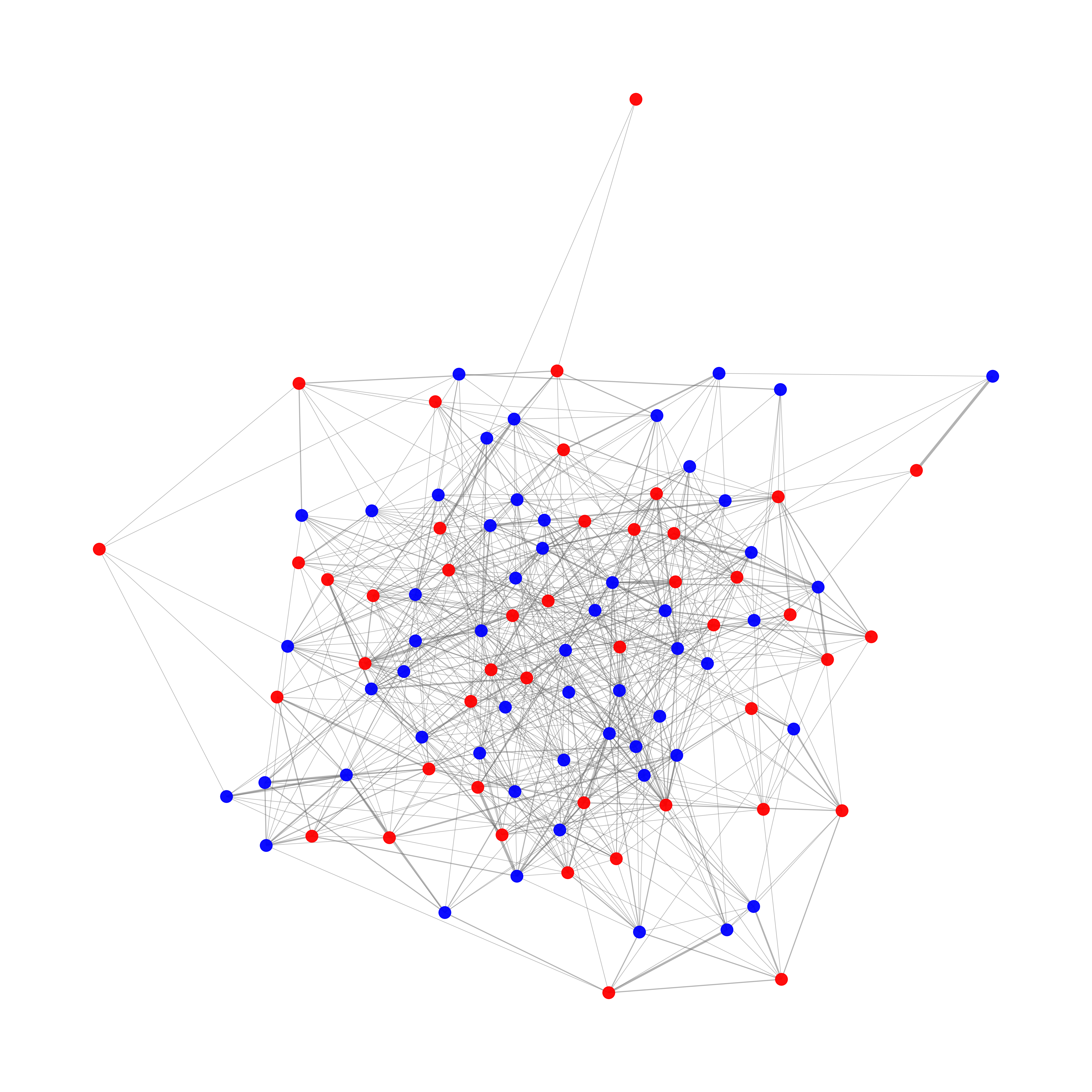}
        \caption{0\% Skill Assertion}
        \label{fig:homophily_sa0}
    \end{subfigure}
 
    \caption{(a) When 100\% of agents assert skills, high assortativity and disconnected components emerge. (b) When 0\% of agents assert skills, a dense, integrated collaborative web forms.}
    \label{fig:homophily}
    \Description{(a) When 100\% of agents assert skills, high assortativity and disconnected components emerge. (b) When 0\% of agents assert skills, a dense, integrated collaborative web forms.}
\end{figure}

Agent traits significantly altered the topology of the emergent collaboration network. High levels of skill assertion led to high network assortativity ($0.578$) and modularity (Figure \ref{fig:homophily_sa1}). Agents naturally formed homophilous clusters, interacting only with those sharing their specialty. This "siloing" effect fragmented the network into disconnected components, increasing the average shortest path length and reducing global integration. As skill assertion decreased, the network transformed from isolated specialist clusters into a dense, integrated collaborative web.

\section{Discussion \& Conclusion}

Micro-level agent traits and task structures interact to shape the success and failure of ad-hoc teamwork. We show that rigid specialization leads to the specialist's dilemma---generating bottlenecks, inequality, and fragmented social silos. We also show that increasing communication costs can be beneficial if it reduces redundant teamwork. These findings suggest that effective MAS design must actively counteract the tendency for specialists to isolate, perhaps by algorithmically encouraging agents to occasionally accept tasks outside their primary expertise to maintain system-wide resilience.


\balance 

\bibliographystyle{ACM-Reference-Format} 
\bibliography{sample, other_sample}


\end{document}